\documentclass{ifacconf}

\usepackage{graphicx}      %
\usepackage{natbib}        %
\usepackage{amsmath} %
\usepackage{amssymb}  %
\usepackage{todonotes}
\usepackage[caption=false]{subfig}
\usepackage{algorithm}
\usepackage{float} 
\usepackage{tikz}
\usetikzlibrary{shapes,arrows}
\usepackage{verbatim}
\usepackage{etoolbox}

\usepackage{placeins} %

\let\classAND\AND
\let\AND\relax
\usepackage{algorithmic}
\let\AND\classAND
\AtBeginEnvironment{algorithmic}{\let\AND\algoAND}

\floatname{algorithm}{Algorithm}
\renewcommand{\algorithmicrequire}{\textbf{Input:}}
\renewcommand{\algorithmicensure}{\textbf{Output:}}

\newcommand{\yref}{y^{\text{ref}}_t}

\begin{document}

\begin{frontmatter}

\title{Aiding reinforcement learning for \\ set point control\thanksref{footnoteinfo}} 
\thanks[footnoteinfo]{
This research was financially supported by the project \emph{NewLEADS - New Directions in Learning Dynamical Systems} (contract number: 621-2016-06079), funded by the Swedish Research Council.}
\author[First]{Ruoqi Zhang}
\author[First]{Per Mattsson}
\author[First]{Torbj\"{o}rn Wigren}

\address[First]{Department of Information Technology, Uppsala University,
   75105 Uppsala, Sweden (e-mail: \{ruoqi.zhang, per.mattsson, torbjorn.wigren\}@it.uu.se).}

\begin{abstract}          

While reinforcement learning has made great improvements, state-of-the-art algorithms can still struggle with seemingly simple set-point feedback control problems. One reason for this is that the learned controller may not be able to excite the system dynamics well enough initially, and therefore it can take a long time to get data that is informative enough to learn for good control. The paper contributes by augmentation of reinforcement learning with a simple guiding feedback controller, for example, a proportional controller. 
The key advantage in set point control is a much improved excitation that improves the convergence properties of the reinforcement learning controller significantly. This can be very important in real-world control where quick and accurate convergence is needed. The proposed method is evaluated with simulation and on a real-world double tank process with promising results.
\end{abstract}


\end{frontmatter}
\section{Introduction}
During the last decade, reinforcement learning (RL) has made significant improvements, resulting in for example computers playing GO \citep{silver2016masteringgo}, Atari games \citep{mnih2015human} and even realistic driving games \citep{wurman2022outracing}, performing beyond human-level. 
This has also led to an increased interest in applying RL to nonlinear control problems. In this paper the focus will be on set point control. In this case (model-free) RL-methods can be seen as (direct) nonlinear adaptive control  methods \citep{krstic1995nonlinear, astrom2008adaptive}, and the same effects that limit adaptive control are likely to affect RL. In this paper it will indeed be seen that even for the relatively simple problem of level control in a cascaded double tank process, state of the art RL-methods may require long times to learn, sometimes they even fail completely. While there are several reasons why RL-methods struggle, this paper argues that one major reason is that they initially provide limited excitation of different amplitudes, and it is shown by experiments that this problem can be mitigated by the introduction of a simple guiding feedback controller.

In system identification and adaptive control it is well-known that in order to learn how to control a system, the data collected must excite the system dynamics. In RL this is called exploration. There is a vast literature on exploration in RL, cf. \citep{pmlr-v119-jin20d_reward_free_Exploration, tang2017exploration, nair2018overcoming_exploration}, but the practical state of the art algorithms for deep RL often just add white probing noise to the input, either explicitly \citep{fujimoto2018TD3, silver2014ddpg} or implicitly via a stochastic control policy \citep{schulman2017proximal, haarnoja2018sac}. While this is an efficient way to excite different frequencies, it does not by itself ensure that the system gets excited in amplitude, even though this is important for nonlinear systems, cf. \citep{wigren1993recursive}.

In adaptive set-point control of nonlinear systems, the amplitude excitation is crucial, since the controller should learn how to handle different reference levels. This control objective is often called multi-goal in RL \citep{plappert2018multigoal}. One problem for RL in such a setting is that it begins the learning without any prior knowledge. Hence, initially the RL-controller uses more or less white noise as input. The data collected in this way will typically not excite amplitudes sufficiently in the whole range of the control objective, therefore it can take a long time for the RL controller to learn even how to move towards the reference level. There is also a risk that the controller ends up in a saturated state, where the white probing noise has no effect on the output, which again leads to insufficient excitation. However, the problem of moving in the general direction of the reference signal is handled by feedback control, and it is often sufficient to use a roughly tuned proportional (P) controller to achieve the purpose. 

This paper thus contributes by augmentation with a prior guiding feedback controller to any standard RL method. This controller ensures that the process react to commanded set point changes from the start, thereby providing the amplitude excitation needed for the RL-method to learn a good control policy. The paper argues that a simple P-controller is often sufficient, as illustrated by simulations and experiments on a real-world process. In other words, a simple feedback controller can ensure that the output will change when the set point changes, thus giving a more efficient exploration of different amplitudes.
Note that the cascaded tank process used in the paper could be controlled with PID control. However, the scope of the paper is {\it not} to advocate RL for set-point control for  mildly nonlinear processes, it is rather to demonstrate that set-point controlling RL methods need improvements also for such mildly nonlinear processes, to perform reasonably well. The proposed enhancements  of  the paper are therefore believed to become  critical when RL
is applied to set point control of severely nonlinear systems.
    
While the proposed guiding feedback controller can be combined with any RL-method, the PPO-method by \cite{schulman2017proximal} is used in this paper. One reason for this is that PPO is constructed so that each adaptive update of the control policy is limited in size, and therefore the risk that the training process immediately forgets about the prior controller is reduced.

There are prior work with similar ideas. In \citep{silver2018residual} and \citep{johannink2019RRL}, prior controllers in the form of ad-hoc algorithms or model-based predictive control in a fixed-goal/set point setting were used. In this paper the focus is on the excitation of amplitudes, and it is argued that even simpler strategies can improve RL-methods substantially. It can also be noted that in these prior works off-policy RL-methods are used, and the performance of the adaptive controller tends to first degrade compared to the prior controller before it starts to improve again. In this paper the experiments indicate that when the prior controller instead is combined with PPO, the performance tends to improve from the first update of the control policy.

The paper is organized with the problem statement in Section~\ref{sec:ps}, followed by a brief review of RL combined with a discussion on the exploration/excitation problem in Section~\ref{sec:multi}. The proposed method is given in Section~\ref{sec:proposed}, and is experimentally evaluated in Section~\ref{sec:exp} and \ref{sec:results}. Finally conclusions are found in Section~\ref{sec:conclusions}.

\section{Problem Statement}
\label{sec:ps}
Consider a discrete-time nonlinear state space model
\begin{align}
\label{eq:sp:model}
x_{t+1} &= f(x_t, u_t, w_t) \\
y_t &= h(x_t)
\end{align}
where $f$ and $h$ are unknown functions, $x_t \in \mathbb{R}^{n_x}$ is the current state of the system, $u_t \in \mathbb{R}^{n_u}$ is the input, $y_t \in \mathbb{R}^{n_y}$ is the output, $w_t \in \mathbb{R}^{n_w}$ is the system noise, and $t$ is the time step. 

The problem considered in the paper is to use RL to train a state feedback set point control policy,
\begin{equation}
u(t) = g(x_t, \yref)
\end{equation}
to make the output $y_t$ track a reference level $\yref$. In order to allow for varying set points, the framework of multi-goal RL will be utilized.

\section{Multi-goal Reinforcement Learning }\label{sec:multi}
In multi-goal RL, the aim is to find a goal-conditioned policy that maximizes the discounted future return 
\begin{equation}\label{eq:Gt}
G_t = \sum_{k=0}^{\infty} \gamma^k r_{t+k+1},
\end{equation}
where $\gamma$ is a discount factor, and $r_{t} = R(x_t, u_t, \yref)$ is a scalar reward (negative cost). In the standard RL-setting, the reward is typically determined by the state $x_t$ and the input $u_t$ alone. This would however restrict the controller to handle a fixed set point, so here $\yref$ is included to allow for arbitrary and varying set points. This setting is sometimes called multi-goal RL~\citep{plappert2018multigoal}, where the set points $\yref$ are the goals.

In model-free RL the policy is trained without explicitly identifying the system dynamics of \eqref{eq:sp:model}, and it is thus related to direct adaptive control \citep{astrom2008adaptive}. 
During training new samples are collected while the current control policy is running on the system, i.e. in closed-loop. This data is then used in order to improve the policy.

\subsection{Exploration}
In system identification and adaptive control, it is well known that the system dynamics must be excited. Two common ways to achieve such exploration are to either use a deterministic control policy and add random probing noise or to directly train a stochastic policy where the variance gives excitation of different frequencies. 

In this paper a stochastic policy will be utilized during training. That is, the policy can be seen as a distribution $\pi_\theta(u_t| x_t, \yref)$, where $\theta$ is the parameter vector to be trained.
For example, the input $u_t$ can be drawn from a Gaussian distribution with mean $g_{\theta}(x_t, \yref)$ and standard deviation $\sigma_\theta$. Alternatively, it can be seen as a deterministic control policy with added probing noise, 
\begin{equation}
u_t = g_{\theta}(x_t, \yref) + \varepsilon_{\theta}
\end{equation}
where $\varepsilon_\theta$ is zero mean probing noise with standard deviation $\sigma_{\theta}$. Note that since the standard deviation depends on $\theta$, it can be adjusted adaptively during training. This is somewhat related to the dual control problem in optimal control \citep{astrom1970introduction, florentin1962optimal}.

While well-chosen zero mean probing noise $\varepsilon_\theta$ can ensure good excitation of different frequencies, the excitation of different amplitudes is also important in training set point controllers for nonlinear systems. In closed-loop system identification, this can be achieved by regularly changing the set point $\yref$ to random levels. However, in RL the problem is that initially the mean $g_\theta(x_t, \yref)$ is typically a random control policy, so it will not necessarily react to changes in the set point, and therefore the collected data does initially not excite different amplitudes sufficiently. In Section~\ref{sec:proposed} a method to overcome this problem is presented.

\subsection{Universal value functions}
An important concept in RL is the state-value function, which is the expected value of $G_t$ in \eqref{eq:Gt}. This shows how rewarding the current state is in terms of expected future returns.
In order to handle the multiple-goal setting, the value function can be extended to the universal value function \citep{schaul_universal_nodate}, which depends not just on the state, but also the set point $\yref$, and is defined as  
\begin{equation}
V_\pi(x_t, \yref) = \mathbb{E}_\pi[G_t \vert x_t, \yref],
\end{equation}
where $\mathbb{E}_\pi[\cdot]$ denotes the expected value while using policy $\pi$ and $t$ is any time step. Similarly, the so called $Q$-value function, estimating the value of a state-input pair, is defined as 
\begin{align}
Q_\pi(x_t, u_t, \yref) = \mathbb{E}_\pi[G_t\vert x_t, u_t, \yref],
\end{align} which indicates the expected value starting from the state $x_t$ with input $u_t$, and after that following the policy $\pi$.
When improving the policy, it is relevant to know how much better a specific input is in terms of a certain state as compared to the input given by the current policy. The advantage function is defined as
\begin{align}
A_\pi(x_t, u_t, \yref) =  Q_\pi(x_t, u_t, \yref) - V_\pi(x_t, \yref),
\end{align} and it corresponds to how much better the input $u_t$ is  than taking an input $u$ from the policy distribution $\pi$.
The value functions $V_\pi$, $Q_\pi$ and $A_\pi$ can be estimated, for example, using Monte Carlo methods or Temporal-Difference Learning algorithms~\citep{sutton2018reinforcement}. There are several choices for the advantage estimator. The estimator used in this paper is the generalized advantage estimator \citep{schulman_high-dimensional_2018} to reduce the variance. 

\subsection{Policy Gradient Methods}
\label{subsection:ppo}
Policy Gradient (PG) algorithms are a class of model-free RL algorithms that directly optimize the policy $\pi_\theta$ with respect to $\theta$, with the goal of finding a policy that maximizes $J(\theta) = V_{\pi_{\theta}}(x_t, \yref)$. According to the policy gradient theorem with advantage functions \citep{sutton2018reinforcement}, the gradient can be computed as
\begin{align}
\nabla J(\theta) &= \mathbb{E}_\pi \left [ 
A_{\pi_\theta}(x_t, u_t, \yref) \nabla  \log \pi_\theta (u_t \vert x_t, \yref)
\right ],
\end{align} where $\mathbb{E}_{\pi_{\theta}} [\cdot]$ indicates the expectation over the state-input pairs distribution under the policy $\pi_\theta$. Thus, a commonly used gradient estimator has the form,
\begin{align}
\hat{g} = \hat{\mathbb{E}}_{\pi_\theta} \left [ 
\hat{A}_{\pi_\theta}(x_t, u_t, \yref) \nabla\log \pi_\theta(u_t \vert x_t, \yref) 
\right ],
\end{align}where $\hat{\mathbb{E}}_\pi[\cdot]$ denotes the empirical average over a batch of samples and $\hat{A}$ is an estimator of the advantage function. In implementations, the gradient can be automatically computed by automatic differentiation software, e.g., PyTorch \citep{NEURIPS2019_pytorch}, through constructing a surrogate objective function, 
\begin{align}
J^{PG}(\theta) = \hat{\mathbb{E}}_{\pi_\theta} \left [ 
\hat{A}_{\pi_\theta}(x_t, u_t, \yref) \log \pi_{\theta} (u_t \vert x_t, \yref) 
\right ] \label{eq:Jpg},
\end{align}whose gradient is the gradient estimator $\hat{g}$. One of the most popular PG-algorithms is the PPO-algorithm \citep{schulman2017proximal}, which clip the criterion \eqref{eq:Jpg}  to limit the step of the policy update. In should be noticed that, in the paper, unlike in PPO, the stochastic policy is only used during the training process.

\section{Proposed method}\label{sec:proposed}

 \tikzstyle{block} = [draw, rectangle, 
     minimum height=3em, minimum width=6em]
 \tikzstyle{sum} = [draw, circle, node distance=1cm]
 \tikzstyle{input} = [coordinate]
 \tikzstyle{output} = [coordinate]
 \tikzstyle{pinstyle} = [pin edge={to-,thin,black}]
 \tikzstyle{branch}=[fill,shape=circle,minimum size=3pt,inner sep=0pt]

 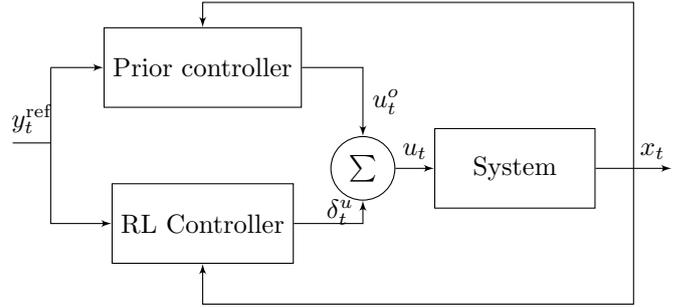
\begin{figure}
     \centering
     \begin{tikzpicture}[auto, node distance=2cm,>=latex']
     \node [input, name=origin]{};
     \node [input, name=input, above right=1.0cm and 0.5cm of origin] {};
     \node [block, right of=input] (priorcontroller) {Prior controller};
     \node [block, below=1cm of priorcontroller] (rlcontroller) {RL Controller};
     \node [sum, below right=0.5cm and 0.5cm of priorcontroller] (sum) {\tiny$\sum$};
     \node [block, right of=sum,
	     node distance=2.0cm] (system) {System};
     \draw [->] (priorcontroller) -| node[near end,name=uo] {$u^o_t$} (sum);
     \draw [->] (rlcontroller) -| node[near end,name=url] {$\delta^u_t$} (sum);
     \draw [->] (sum) -- node[name=u] {$u_t$} (system);
     \node [output, right=0.5cm of system] (output) {};
     \node [output, right=0.5cm of output] (output2) {};

     \draw [-] (origin) -- node {$\yref$}++(0.5,0);
     \draw [->] (input) -- node {} (priorcontroller);

     \draw [->](input) (input.south) |- (rlcontroller);
     \draw [-] (system) -- (output);
     \draw [->] (output) -- node[name=y]{$x_t$} (output2);
     \draw [->] (output) |- ++(0,2.2) -|
     (priorcontroller.north);
     \draw [->] (output) |-++(0,-1.8)-| (rlcontroller.south);

 	\end{tikzpicture}
     \caption{Block diagram of the closed loop system using the proposed method.}
     \label{fig:blockdiag}
 \end{figure}
In standard RL, it is typical to start the training from an initial random control policy. This means that initially the controller does not react in a systematic way to changes in the set point, and therefore the collected data may not not excite the system well in amplitude, even if random set points are used.

However, even with limited knowledge about the system it is often possible to design a controller that at least will change the output towards the reference level after it has changed.  Even a proportional feedback controller 
\begin{equation}
    g^o(x_t, \yref) = K(\yref - y_t)
\end{equation}
can typically achieve this.

Hence, the idea here is that instead of starting the learning from a random control policy, the controller is given by 
\begin{equation}\label{eq:proposed}
g(x_t,\yref) = g^o(x_t, \yref) + g_\theta(x_t,\yref),
\end{equation}
where $g^o(x_t, \yref)$ is a prior controller, and $g_\theta$ is a residual control policy given by a neural network that is trained using a standard RL-method. 
A block diagram of the closed-loop system is shown in Figure~\ref{fig:blockdiag}.
To start from the prior controller, the policy neural network for $g_\theta$ is initialized with the last layer set to zero. 

There are at least two reasons why it is more efficient to train the residual policy instead of starting from scratch. First of all, if the prior policy is good, then the RL-method just has to make minor improvements. However, the main point here is that a prior controller that nudges the system in the right direction can improve the performance of the RL-methods substantially by ensuring that more exciting data is collected from the start. Also note that the structure in \eqref{eq:proposed} allows the RL-method to eventually cancel out the prior controller if the data indicate that it is not useful beyond the initial data collecting phase. This reduces the risk of locking the RL-method into an inefficient control structure.

\subsection{Algorithm Summary}
\label{sec:algo}
The proposed method is summarized by Algorithm \ref{algo:RRL}. The algorithm is written as on-policy learning, meaning that the policy updates are based on data collected while using the current policy. For each policy update, data is collected using $M$ different set points, where $M>1$, in order to stabilize the training process. The method is also possible to combine with off-policy RL methods, where also data collected while using older polices can be utilized in each update. 

The PPO algorithm is chosen in the paper to optimize the residual policy, this will be denoted as ResidualPPO throughout the rest of the paper.
The clipped surrogate objective in PPO, mentioned in Section \ref{subsection:ppo}, improves the training when using a prior policy to track various reference signals. For each update, the training data only contains a limited number of reference signal samples and thus it may be better not to move away from the old policy too fast. 
\begin{algorithm}
\caption{Residual Multi-goal Reinforcement Learning }
\begin{algorithmic}[1]
\label{algo:RRL}
\renewcommand{\algorithmicrequire}{\textbf{Input:}}
\renewcommand{\algorithmicensure}{\textbf{Output:}}
\REQUIRE Policy $\pi_\theta$, prior controller $g^o$, goal space $\mathcal{G}$
\FOR {iteration=$1,2,\dots$}
\STATE Start collecting data with $M$ different goals
\FOR{$m=1,2,\dots,M$}
\STATE Sample $y^{\text{ref}}_m$ from $\mathcal{G}$ uniformly
\STATE Sample initial state of dynamic system $x_0$ (for real-world systems, just start from whatever state the system is in)
\FOR{$t=0\dots T-1$,}
\STATE Get prior input  
$u^o_t=g^o(x_t, y^{\text{ref}}_m)$
\STATE Sample input from policy distribution $\pi_\theta$ by $\delta^u_t \sim \mathcal{N}(g_\theta(x_t, y^{\text{ref}}_m), \sigma_\theta(x_t, y^{\text{ref}}_m))$
\STATE Apply the input
$u_t = u^o_t +\delta^u_t$
\STATE Get next state $x_t$ from dynamics and reward signal $r_t$ from reward function 
\STATE Store $(x_t, y^{\text{ref}}_m, u_t, x_{t+1}, r_t)$
\ENDFOR 
\ENDFOR
\STATE Optimize $\theta$ using RL algorithm, e.g., PPO.
\ENDFOR
\end{algorithmic} 
\end{algorithm}

\section{Experiments}
\label{sec:exp}
The above ideas were evaluated on a cascaded double tank system, both in simulation and in real-world. The reward function is the negative value of the absolute distance to the reference signal and the reference signal is generated randomly at the beginning of the episode.
As for the evaluation, due to the random set points for each episode, Monte Carlo simulations are performed with 100 runs to compute the expected performance. 
A modified version of Stable-Baselines3 \citep{raffin2019stable} was used for the implementation. 
\subsubsection{Water Level Control with Simulation}
\begin{figure}
\centering
\includegraphics[width=0.6\linewidth]{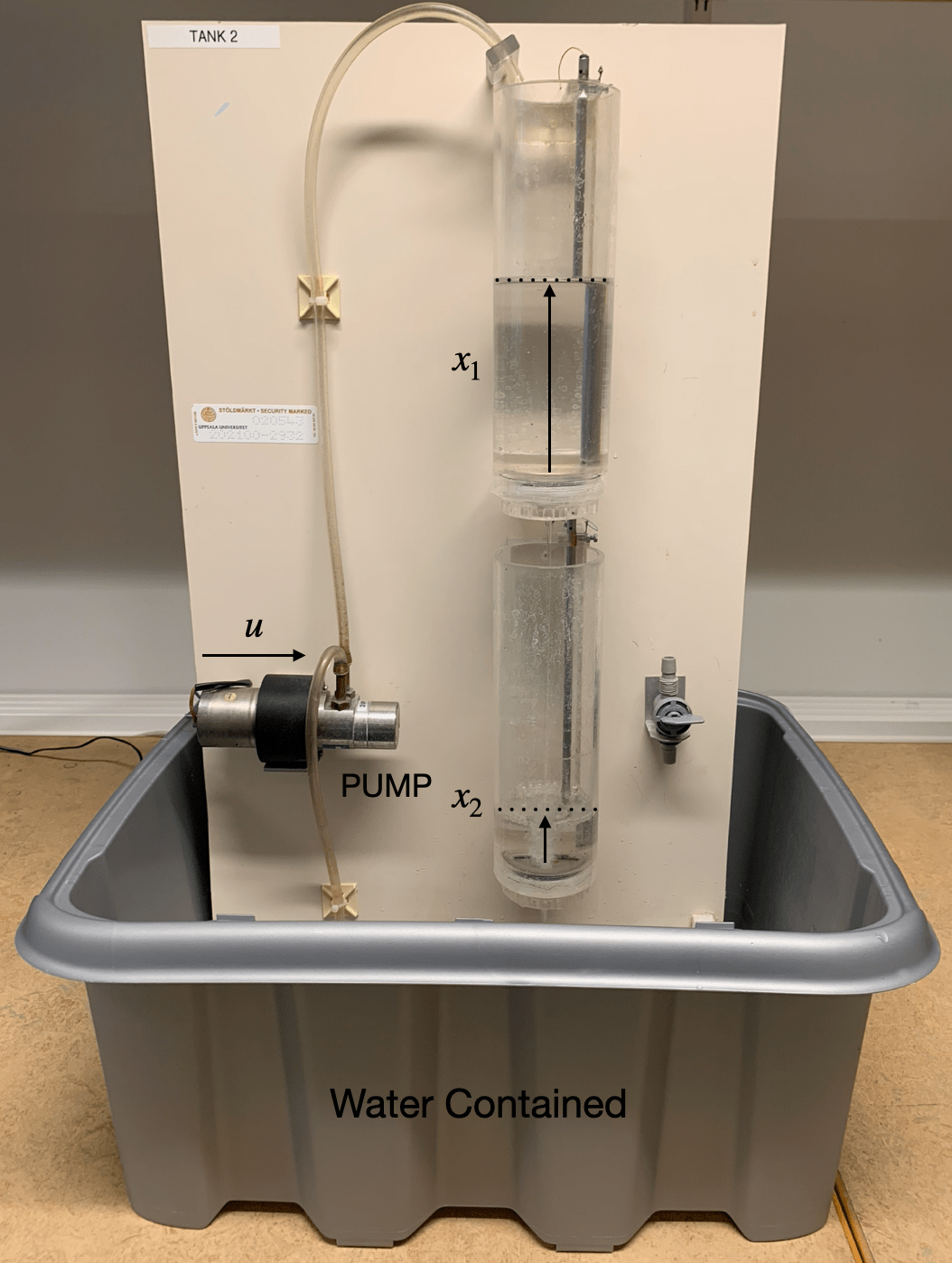}\caption{The water tank process.}
\label{fig:watertank-process}
\end{figure}

The water tank system consists of two identical tanks mounted on top of each other.
As depicted in Fig. \ref{fig:watertank-process}, the water flows into the upper tank with the flow controlled by an electrical pump. There is a small hole in the bottom of each tank, so the water can flow from the upper tank to the lower tank and from the lower tank to the container below the tank system. The input signal $u_t$ is the voltage applied to the pump and the goal is to keep the water at a certain level in the second tank. 

Belowe the notation that $x(t)$ is the signal at continuous time $t$, while $x_t$ is the signal at the $t$th discrete time-step is used, and for the example the sampling period is set to 2 seconds. 

The state of the system contains the water levels of the two tanks, 
\begin{align}
x(t) = \begin{bmatrix}x_1(t) & x_2(t)\end{bmatrix}^T
\end{align} and the output is
\begin{align}
y(t) = x_2(t).
\end{align}
A standard model \citep{wigren2013_watertank} for this system is given by
\begin{align}
\label{eq:modelwatertank}
\begin{split}
\dot{x}_1(t) &= -\frac{a_1}{A_1}\sqrt{2gx_1(t)} + \frac{K_\text{pump}}{A_1}V_p(t),\\
\dot{x}_2(t) &= \frac{a_1}{A_2}\sqrt{2gx_1(t)} - \frac{a_2}{A_2}\sqrt{2gx_2(t)},
\end{split}
\end{align} where $K_\text{pump}$ is the pump constant, $V_p$ is the voltage, $a_1$ and $a_2$ are the areas of the holes, $A_1$ and $A_2$ are the areas of the cross sections of the tanks.

The values of the parameters of the model are the measured parameters from the real-world water tank with the ratio of the area of holes and the area of the cross sections of the tank being $a/A=0.0019$, and the range of input voltage from the pump being $[0,10]$ V. The pump constant is $0.12$ cm/Vs. The initial levels of the two tanks and the reference signal are generated uniformly between $0$ and $10$~cm at the beginning of each episode, which by itself gives some extra amplitude excitation. The reference signal was changed every 400 s. 
The prior policy for this task is a discrete-time proportional controller,
\begin{align}
    u^o_t = g^o(x_t, \yref) = -K_p (\yref-y_{t}) + u_0
\end{align}
with $K_p = 2$ and $ u_0=5$. Note that this is not a well-tuned P controller. The purpose of these experiments is not to compare the use of a P controller with RL methods, the purpose is instead to show the the inclusion of a roughly tuned prior P controller can improve standard RL methods substantially.

\subsubsection{Real-world Water Tank Level Control}
The setup for the real water tank is the same as for the simulated one, with a sampling period of 2 seconds and a prior P-controller.  However, the system is a bit different since the real tank can overflow. In addition, when the upper tank overflows, parts of the water enter the lower tank, and the rest flows to the container. This is not taken into account in simulations. Also, instead of using measured water levels as output, the sensors give noisy voltage measurements instead of the true water levels in each tank. The voltages of the sensors are proportional to the water levels.

\begin{figure*}[htpb!]
  \centering
  \subfloat[Training performance comparison between PPO method and ResidualPPO in simulation with $10$ different seeds. The methods are evaluated with 100 different set points that are drawn uniformly from $0$ to $10$. \textit{Orange} corresponds to the average performance of the P-controller which is the prior controller of ResidualPPO. \label{fig:res-watertank}]{%
    \includegraphics[width=0.4\linewidth]{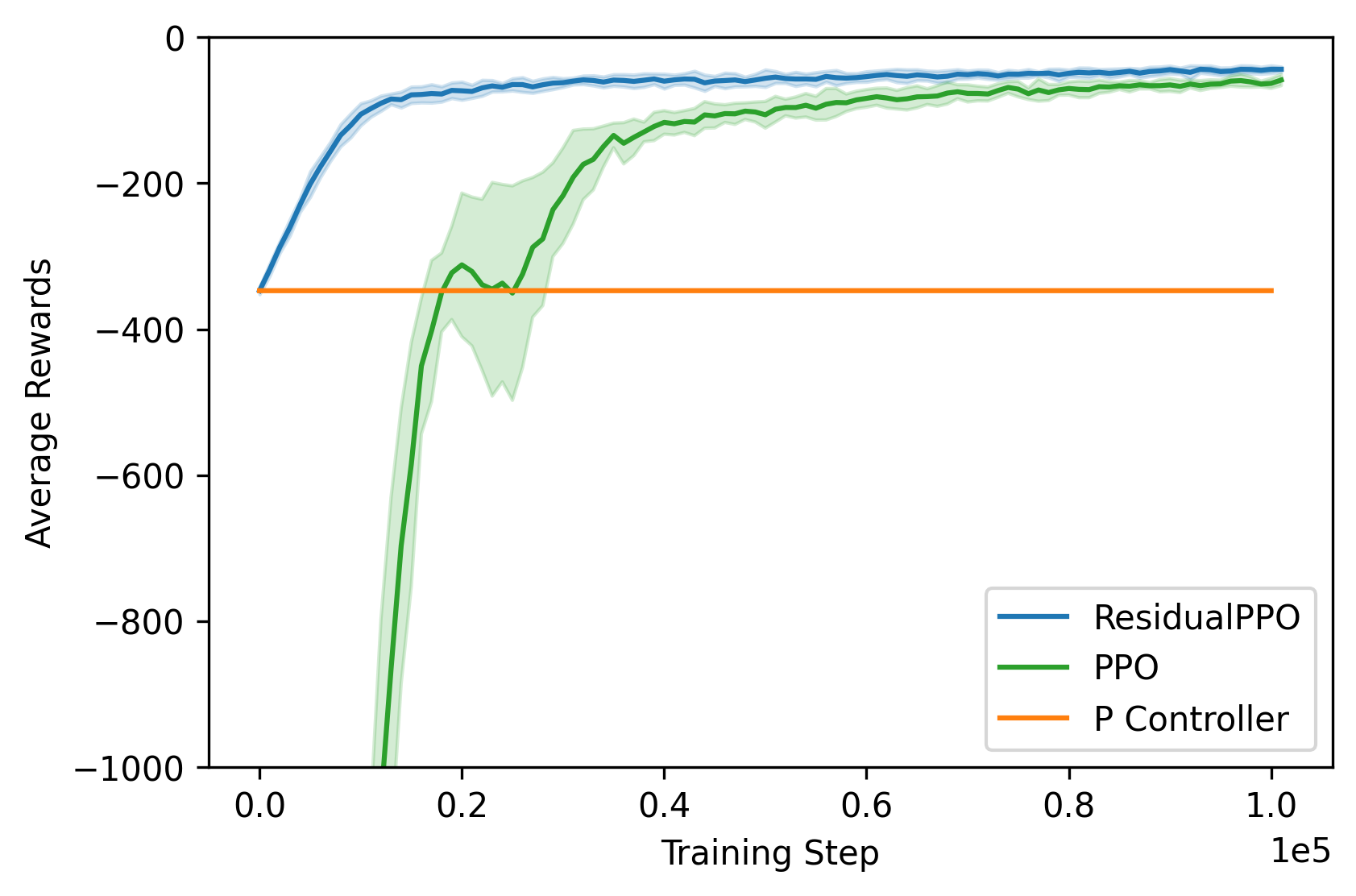}%
  }
  \hfill
  \subfloat[Simulation comparison between PPO and Residual PPO with 5 different set points (\textit{black dash line}). The methods are evaluated 200 steps for each reference signal. The $y$ axis represents the level in the second tank and \textit{orange} corresponds to the performance of the prior P-controller used in ResidualPPO. \label{fig:test-watertank}]{%
    \includegraphics[width=0.4\linewidth]{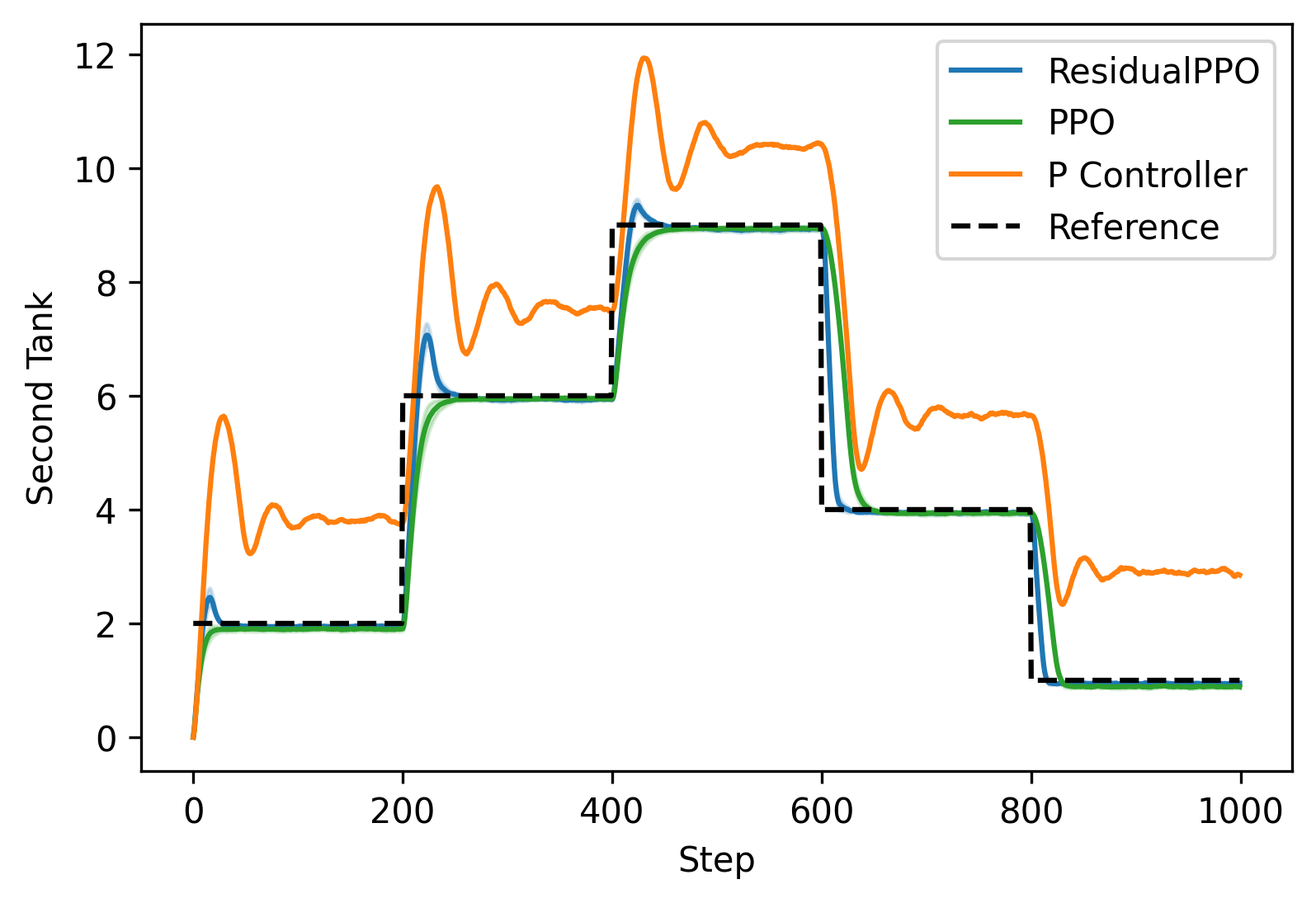}%
  }

  \centering
  \subfloat[Real-world comparison between ResidualPPO with different training steps (5k, 10k, 15k) on  5 different set points (\textit{black dash line}). The $y$ axis represents the level in second tank and \textit{red} corresponds the performance of prior P-controller used in ResidualPPO.\label{fig:real-res}]{%
    \includegraphics[width=0.4\linewidth]{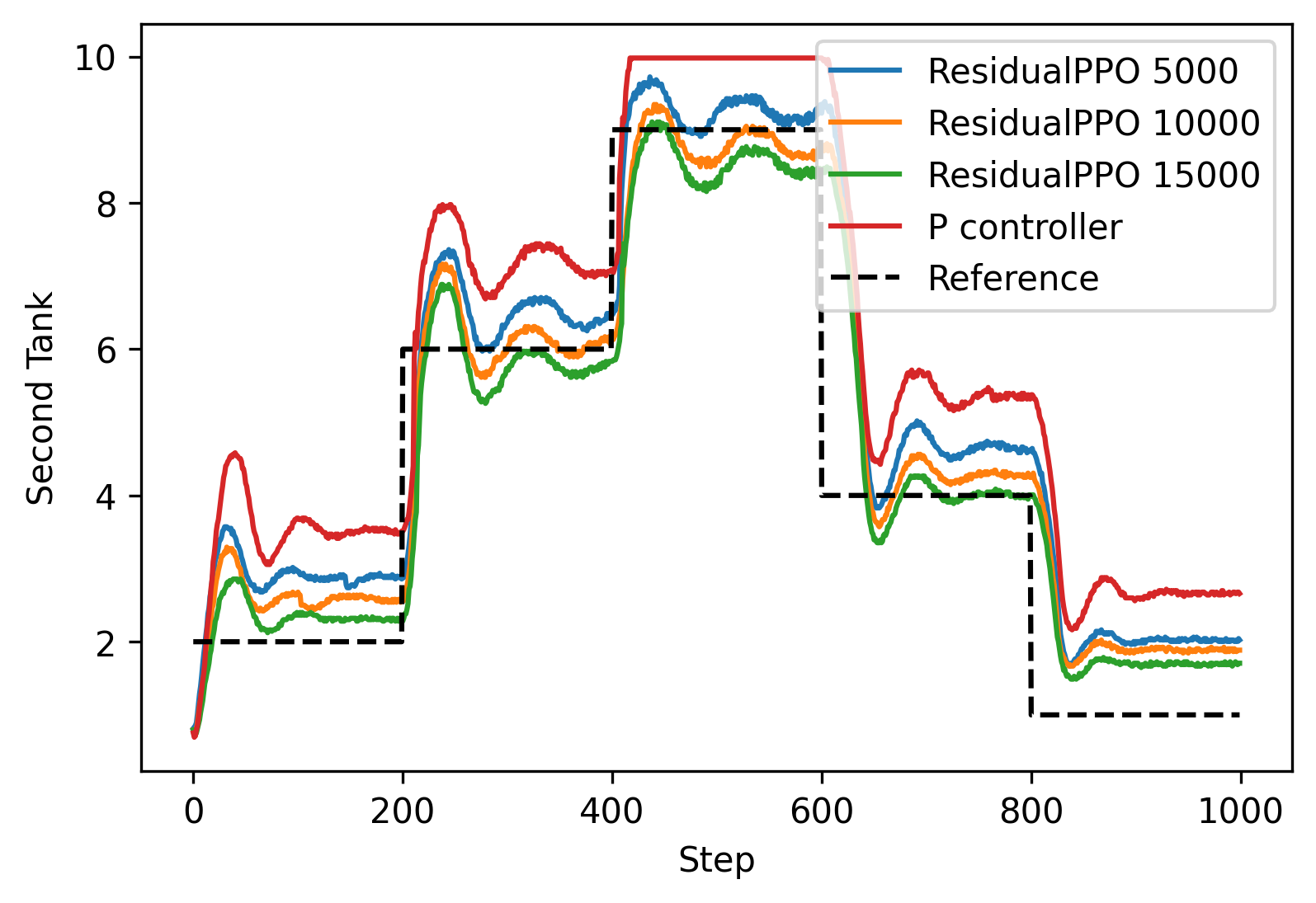}%
  }
  \hfill
  \subfloat[Real-world comparison between PPO with different training steps (5k, 10k, 15k) on  5 different set points (\textit{black dash line}).  The $y$ axis represents the level in the second tank and \textit{red} corresponds to the performance of the prior P-controller used in ResidualPPO. Note that \textit{blue}, \textit{orange} and \textit{green} lines overlap each other in the plot.]{%
    \includegraphics[width=0.4\linewidth]{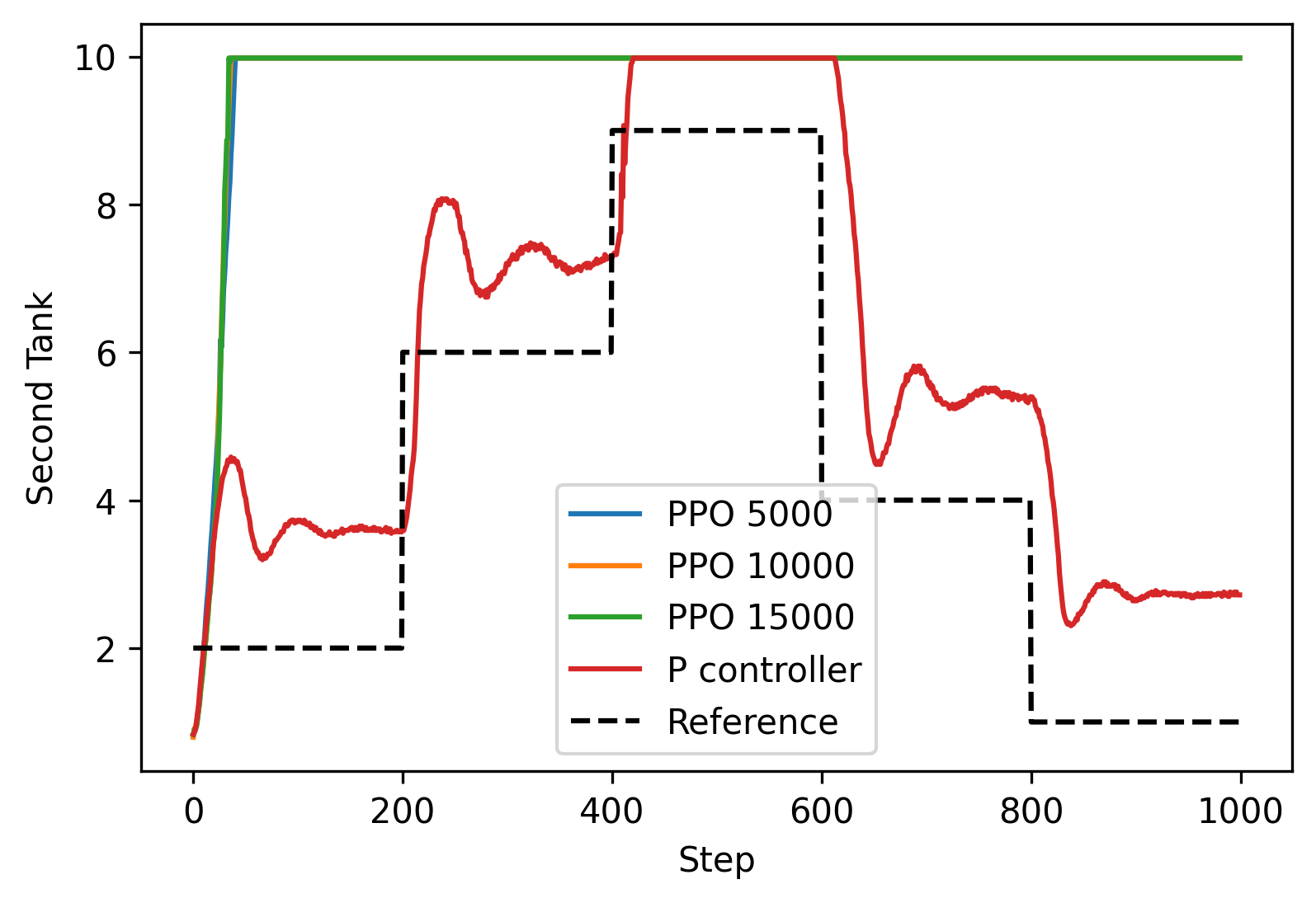}%
  }
  \caption{Experiment results on simulated (a,b) and real-world (c,d) water level control.}
  \label{fig:realtank}
\end{figure*}

\subsection{Policy Network and Training Details}
For consistency, the same actor-critic architecture is used for PPO agents with or without a prior controller. 
The network for the residual policy consists of three fully connected layers of 128 units each with a tangent function as a nonlinear activation function except for the last layer.
In all tasks, the inputs are normalized to lie between -1 and 1 when the networks are trained, in order to stabilize the training.
The agents are optimized with minibatch Adam \citep{kingma2014adam} with  minibatch size $N_\text{batch} = 256$ for $K=16$ epochs.
The learning rate is set to $3e^{-4}$.

For each update, the training samples contain $M=5$ different reference signals. 
The hyperparameters of the PPO algorithm are tuned following the recent study on Deep RL reproducibility by \cite{andrychowicz2021matters}, with GAE parameter $\lambda = 0.97$, discount factor $\gamma = 0.99$, clipping parameter $\epsilon=0.2 $ and coefficients in complete objective $c_1=1.0, c_2=0.02$.

The agents are evaluated 10 times, with seeds from 1 to 10.
Unlike in \citep{silver2018residual}, where the value function estimator is trained alone for a "burn in" period while leaving the policy fixed, the value function estimator starts from scratch.

\section{Results}\label{sec:results}
In this section, the results for PPO are compared with ResidualPPO, where PPO is augmented with a prior P-controller as proposed in Section~\ref{sec:proposed}.
All empirical results are presented with mean and standard deviation across $10$ random seeds.
It is again noted that the cascaded tank process can be well controlled e.g. with PID control. Thus, it is {\it stressed} that the purpose here is to demonstrate the need for aiding control when RL is used.

\subsection{Water Level Control with Simulation}
    Both PPO and ResidualPPO achieve the ability to deal with varying level control, but ResidualPPO requires fewer training samples and shows a smaller training variance. As shown in Fig. \ref{fig:res-watertank}, the PPO algorithm needs $20000$ to $25000$ samples to reach the P-controller's performance, which is the initial performance of the  ResidualPPO. 
    
    Note that although the value function approximator is initialized with poor performance with no pre-train period, there is no drop in performance of the ResidualPPO during training unlike in \citep{silver2018residual}. One explanation may be that the PPO is an on-policy algorithm and the agent is trained using 5 different reference levels for each update. Also, as mentioned in Section~\ref{sec:algo}, the clipped surrogate objective in PPO avoids the large step between a new policy and the old one.
    It can be seen that from Fig.~\ref{fig:test-watertank} that both PPO and PPO with a P-controller perform well reaching the reference levels, $[2,6,9,4,1]$. It may be noted that ResidualPPO has a small overshoot, but this actually results in a slightly higher reward than for the PPO.

\subsection{Real-world Water Tank Level Control}
 As shown in Fig \ref{fig:realtank}, ResidualPPO and PPO with different training samples are tested to track different reference levels in the second tank. The ResidualPPO improves the performance, and with more training samples it is able to track various reference signals at a smaller cost. By contrast, PPO without any prior controller fails this task when using $15000$ training samples. 
 
 It can be noted that PPO seems to find a policy that always maximizes the input after the first 5000 samples. From this controller, the new data that is collected is noninformative regarding the exciting system dynamics. Even though white probing noise is added during training, it is in a saturated state of the system, so the output will not change significantly. The residual version however will get data that excites different amplitudes immediately, and can thus learn a better policy faster.

\section{Discussion and Conclusion}\label{sec:conclusions}
The paper argues that when RL-methods are applied to set point control problems, the inclusion of a conventional feedback controller can improve the learning of a control policy. The numerical experiment also shows that with the prior policy the RL-method requires fewer training samples and gives a more stable training performance than the standard RL-method. One reason is that the prior policy gives better amplitude excitation for changing set points, and thus more relevant training data. 

When applied to real world data the aiding controller enables the RL controller to converge to a well performing closed loop system, while un-aided  RL controller gets stuck in a saturated state.

Interesting topics for further research include stability analysis of control loops employing the proposed methods. The basic idea can also be expanded in several directions, deriving other adaptation schemes. It would also be interesting to compare the performance of the proposed RL method, to classical nonlinear control schemes, from nonlinear optimal control or the geometric field of nonlinear control.

\newpage
\bibliography{RL.bib}

\end{document}